\documentclass{amsproc}

\newtheorem{theorem}{Theorem}[section]

\theoremstyle{definition}

\theoremstyle{remark}

\def\av{{\rm av}}
\def\half{{\textstyle \frac{1}{2}}}
\def\ds{\displaystyle}

\numberwithin{equation}{section}

\begin{document}

\title
[One Dimensional Regularizations  of the  Coulomb Potential]
{One Dimensional Regularizations  of the  Coulomb Potential
 \\ with Application to Atoms in Strong Magnetic Fields}


\thanks{\copyright 1999 by M.B. Ruskai. Reproduction of this article,
  in its entirety, by any means is permitted for non-commercial
  purposes.}

\author{Raymond Brummelhuis }
\address{Department of Mathematics \\  Universit\'e de Reims, 
\\ F-51687 Reims, France}
    \email {raymond.brummelhuis@univ-reims.fr}

\author{Mary Beth Ruskai}
\address{Department of
        Mathematics,   University of Massachusetts  Lowell,   Lowell,
        MA  01854 USA} \email {bruskai@cs.uml.edu} 
\thanks{The second author was supported in part by National Science
        Foundation Grant DMS-97-06981}

\author{Elisabeth Werner }
\address{Department of Mathematics,  
    Case Western Reserve University, 
    Cleveland, OH 44106  USA}
    \email {emw2@po.cwru.edu}


\date{\today}


\begin{abstract}
It is well-known that the functions
${{V_m(x) =  \frac{1}{\Gamma(m+1)} \int_0^\infty
          \frac{u^m e^{-u} }{\sqrt{x^2 + u}} du}}$
arise naturally in the study of atoms in strong magnetic fields
where they can be regarded as one-dimensional regularizations of
the Coulomb potential. 
For many-electron atoms consideration of the Pauli principle
requires convex combinations of such potentials and 
interactions of the form 
$\frac{1}{\sqrt{2}} V_m ( \frac{|x_1 - x_2|}{\sqrt{2}} )$.
We summarize the results of a comprehensive study 
 of these functions including recursion relations,
tight bounds, convexity properties, and connections with
confluent hypergeometric functions.  We also report briefly on
their application in one-dimensional models of 
many-electrons atoms in strong magnetic fields.

\end{abstract}

\maketitle

\section{Introduction}

It is well-known that systems in strong magnetic fields behave 
like systems in one-dimension, i.e., a strong magnetic field  
confines the particles to Landau orbits orthogonal to the
field, leaving only their behavior in the direction of the field
subject to significant influence by a static potential.
Motivated by this general principle and the work of Lieb,
Solovej and Yngvason \cite{LSY} on atoms in extremely 
strong magnetic fields, Brummelhuis and Ruskai \cite{BR1} 
initiated a study of models of atoms in homogeneous strong magnetic
fields in which the 3-dimensional wave-function  has the form
\begin{equation} \label{eq:psi.lowland}
  \Psi({\bf r}_1, {\bf r}_2 \ldots {\bf r}_n) = 
      \psi(x_1 \ldots x_n) \, \Upsilon(y_1,z_1, y_2,z_2, \ldots y_n,z_n)
\end{equation}
where $\Upsilon$ lies in the projection onto the lowest Landau
band for an N-electron system.  We follow the somewhat non-standard
convention of choosing the magnetic field in the x-direction,
i.e, ${\bf B} = (B, 0, 0)$ where $B$ is a constant denoting
the fields strength, in order to avoid notational confusion
with the nuclear charge $Z$.

Such models lead naturally to one dimensional regularizations
of the Coulomb potentials of the form
\begin{eqnarray}  
 \label{vmdef} V_m(x)  & = &
      \int_0^{2 \pi}\int_0^\infty
   \frac{|\gamma_m(r,\theta)|^2 }{\sqrt{x^2 + r^2}} r \, dr \, d\theta \\
     & =&  \frac{1}{\Gamma(m+1)} \int_0^\infty
          \frac{u^m e^{-u} }{\sqrt{x^2 + u}} du \label{vmequiv2} \\
   & = &\frac{2e^{x^2}}{\Gamma(m+1)}
            \int_{|x|}^\infty  (t^2-x^2)^me^{-t^2} dt \label{vmequiv1}
\end{eqnarray}
where  $\gamma_m(r,\theta) = \frac{1}{\sqrt{ \pi m!}} e^{-im\theta} r^m
e^{-r^2/2}$. Recognition that such potentials are important goes back at 
least to Schiff and Snyder \cite{SS} in 1939 and played an
important role in the Avron, Herbst and Simon study \cite{AHS} of
hydrogen.  Recently, Ruskai and Werner \cite{RW} undertook a
detailed study of these potentials, proving
the important property of convexity of $1/V_m$ as well as a
number of other useful properties.  The primary  purpose
of this note is to give a summary of these results in Sections
\ref{sect:Vm.props} and  \ref{sect:Vm.recurs}.  
Before doing that, we briefly discuss
one-dimensional models of atoms in strong magnetic fields
in Section \ref{sect:mag.model} and their implications for the
maximum negative ionization problem in Section \ref{sect:max.neg}.

\section{Atoms in Strong Magnetic Fields}\label{sect:mag.model}
The Hamiltonian for an $N$ electron atom in a magnetic field ${\bf B}$ is 
\begin{equation} \label{eq:Ham.3dim}
H(N,Z,B) = \sum_{j=1}^N \left[ |{\bf P}_j + {\bf A}|^2  
     -\frac{Z}{|{\bf r}_j| } \right] +
     \sum_{j<k} \frac{1}{|{\bf r}_j - {\bf r}_k |} 
\end{equation}
where  ${\bf A}$ is a vector potential such that   
  ${\bf \nabla}  \times  {\bf A } = {\bf B }$. The ground-state energy of   
$H(N, Z, B ) $ is given by   
\begin{equation}
E_0(N,Z,B) = \inf_{\| \Psi \| = 1} 
  \langle H(N, Z, B ) \Psi , \Psi \rangle 
\end{equation}
Let $E_0^{\rm conf}(N,Z,B)$ denote the corresponding minimum
restricted to functions of the form (\ref{eq:psi.lowland}).
For extremely strong fields, it was shown in \cite{LSY} 
 that $E_0 / E_0^{\rm conf} \rightarrow 1$ as 
$B/Z^{4/3} \rightarrow \infty$ with $N/Z$ fixed.

In  \cite{BR1}and \cite{BR2} we consider two special cases
of (\ref{eq:psi.lowland}).  We write the Landau state with angular
momentum $m$ in the x-direction in the form
\begin{equation}
\gamma_m^B(y, z ) =   \frac{B^{(m+1)/2}}{\sqrt{\pi m!}} \, 
  \overline{\zeta }^m e^{-B |\zeta |^2 /2}
\end{equation} where $\zeta = y + i z$.  Then our two special
cases can be described as follows:
\begin{description}
\item[Zero model] In this case, we make the
extremely simple assumption that $\Upsilon$ is simply a product of 
Landau states with $m = 0$, i.e., 
 $\Upsilon =\prod_{k=1}^N  \gamma_{0}(y_k,z_k)$.
\item[Slater model] In this case we assume that $\Upsilon$ is an
antisymmetrized product of Landau states with $m = 0, 1, \ldots N-1$,
i.e.,  $\Upsilon = \frac{1}{\sqrt{N!}} 
[\gamma_{0} \wedge \gamma_{1} \wedge \ldots \wedge \gamma_{N-1} ]$.
\end{description}
Although the first model is somewhat unrealistic, its simplicity
makes it amenable to detailed analysis which yields insight 
into the general situation.  The second model corresponds to the
physically reasonable assumption that $\Upsilon$ is a Slater
determinant.  In this case, the required antisymmetry of the wave 
function is inherent in our assumptions on $\Upsilon$ and the one
dimensional function $\psi$ is symmetric, i.e., the electrons
behave like bosons in one dimension.

It is straightforward to show that
\begin{equation}
  E_0^{\rm conf}(N,Z,B) = \sqrt{B} \inf_{\| \psi \| = 1} 
    \langle h(N, Z, M ) \psi , \psi \rangle  + N B
\end{equation}
where have scaled out the field strength $B$ so that
\begin{equation}\label{eq:1dimHam}
h(N,Z,M) = 
\sum_{j=1}^N \left[
   - \frac{1}{M} \frac{d^2}{dx_j^2} -Z \tilde{V}(x_j) \right] 
  +  \sum_{j< k} \tilde{W}(x_j- x_k)
\end{equation}
and the only remnant of the magnetic field is in the ``mass''  
$M = B^{-1/2}$ and the effective one-dimensional potentials
$\tilde{V}$ and $\tilde{W}$  will be defined below for each model.  

 For the zero model, one easily finds \cite{BR1} that  
\begin{equation} \label{eq:zero}
 \tilde{V}(x) = V_0(x)  \hspace{0.6cm} \hbox{and} \hspace{0.6cm}
\tilde{W}(u-v) = \frac{1}{\sqrt{2}} 
              V_0\left(\frac{|u-v|}{\sqrt{2}}\right).
\end{equation}
Note $V_0(x) \approx \frac{1}{|x|}$ for large $|x|$. Thus, 
for large separations,
\begin{equation}\label{eq:largesep}
  \tilde{W}(u-v) \approx  \frac{1}{\sqrt{2}}
       \frac{\sqrt{2}}{|u-v|} = \frac{1}{|u-v|}.
\end{equation}
However, not only is the singularity removed at $u = v$,
$\tilde{W}(0)$ is smaller by a factor of $\frac{1}{\sqrt{2}}$
than $\tilde{V}(0)$.  This means that if two electrons are
simultaneously near the nucleus, the price paid from the 
electron-electron repulsion is smaller than that gained from
the electron-nuclear attraction.  Although this effect seems to play
an important role in binding additional electrons, it may be partially
offset by the price paid in kinetic energy if one attempts to
constrain both electrons near the nucleus.  See \cite{BR1} for
further discussion.

For the Slater model it can be shown \cite{BR2} that
\begin{equation} \label{eq:V.slater}
 \tilde{V}(x) \equiv V_{\av}^N(x) = \frac{1}{N}\sum_{m=0}^{N-1} V_m(x)
   \hspace{0.6cm} \hbox{and}
\end{equation}
\begin{equation} \label{eq:W.slater}
 \tilde{W}(u-v) = \frac{1}{\sqrt{2}} 
              \sum_{j=0}^{N-1} c_j V_{2j+1}
         \left(\frac{|u-v|}{\sqrt{2}}\right)
\end{equation}
where $c_j > 0~ \forall \, j$ and $\sum_j c_j = 1$ so that the effective
interaction is a convex combination of $V_m$ with odd $m = 1, 3, \ldots
2N-1$, albeit with the same $\frac{1}{\sqrt{2}} $ scaling as in 
(\ref{eq:largesep}).   Note that  the  convex sum in (\ref{eq:W.slater})
above includes contributions from $V_m$ with $m > N$.  Properties (b) and
(d)  of Section \ref{sect:Vm.props}  imply
that $V_m(0)$ is decreasing in $m$.  Therefore, one expects a
decrease in the electron-electron repulsion $ \tilde{W}$ in addition to
that from the factor of $\frac{1}{\sqrt{2}} $.  However, delicate
combinatorics would be needed to verify this exactly.

Some obvious variations on these models are possible and 
discussed briefly in \cite{BR1,BR2}.  It is interesting to note
that if $\Upsilon =\prod_{k=1}^N  \gamma_{m}(y_k,z_k)$ with
$m$ odd, then the convex sum analogous to (\ref{eq:W.slater})
contains only terms $V_{2j}$ with even subscript.

It is also worth noting  that
\begin{equation}  \label{eq:delta}
\lim_{ \beta \rightarrow \infty} 
   \frac{\beta }{\log \beta  } V_m \left( \beta x\right)
 =  \delta (x)   
\end{equation}  
in the sense of tempered distributions.  This implies that the
potentials $\tilde{V}$ and $\tilde{W}$ which occur in our models
have an analogous delta potential behavior as 
$\beta \rightarrow \infty$. The proof \cite{BD} of (\ref{eq:delta})
uses the Fourier transform  (property (k) in  Section 
\ref{sect:Vm.props}) of the  potentials $V_m(x)$, particularly the
observation that
$\widehat{V}_m(\xi)$ has a logarithmic singularity at $\xi = 0$.   
Similar  limiting  behavior (with $\beta = \sqrt{B}$) was observed
  by \cite{LSY} for potentials in the three-dimensional
  Hamiltonian  (\ref{eq:Ham.3dim}).  It can also be shown \cite{BD}
 that if (\ref{eq:1dimHam}) is appropriately rescaled and the
potentials replaced by the corresponding
delta potentials, the result is  a one-dimensional Hamiltonian whose
semi-classical limit (on bosonic wave functions) as
$Z \rightarrow \infty$ is precisely that given by the (fermionic)
hyperstrong functional in \cite{LSY}.  Since their functional
was shown  to describe the $Z, \, B/Z^3 \rightarrow
\infty$ limit of the three-dimensional Hamiltonian (\ref{eq:Ham.3dim}),
this provides additional justification for our models.  Even
the simple-minded zero model has the correct asymptotic behavior.

\section{Maximum Negative Ionization}\label {sect:max.neg}

In the absence of a magnetic field, one expects that the maximum
number of electrons a nucleus with charge $Z$ can bind is
$N_{\max}(Z) = Z + 1$ or $Z+2$.  However, only the somewhat weaker
result of asymptotic neutrality has been proved rigorously
\cite {LSST}.   If electrons behave like bosons, asymptotic
neutrality does not hold and $N_{\max}$ behaves asymptotically
roughly like $1.21Z$.  (See \cite{Sol} for  details and
references to earlier work on bosonic atoms.) 
In \cite{Lb} Lieb gave a simple
argument which showed that $N_{\max} < 2Z + 1$, independent of
particle statistics.  Thus, it may seem somewhat surprising that
\cite{LSY} showed that for atoms in extremely strong magnetic fields
\begin{equation}
  \liminf_{Z, \, B/Z^3 \rightarrow \infty} \frac{N_{\max}(Z)}{Z} \geq 2.
\end{equation}
 The study of one-dimensional models in
\cite{BR1} was initiated, in part, by the hope of proving an
asymptotic upper bound of the form $N_{\max} \leq 2Z$ as
$B, Z \rightarrow \infty$.  Although we did not succeed in
proving such a bound, even for our simplified one-dimensional
models, we believe that they offer considerable insight into
both the reasons for binding an ``extra'' $Z$ electrons and the
reasons why the localization techniques developed to bound
$N_{\max}$ fail in the strong field case.

It is generally believed that enhanced binding occurs in strong
magnetic fields because the field confines the electrons in
two dimensions and effectively reduces the atom to a 
one-dimensional system.  Although there is some truth to
this, it was shown in \cite{LSY} that atoms do not become
truly one-dimensional unless $B > Z^3$ and the field strength
is greater than anything seen on earth.  (Sufficiently strong
magnetic fields do exist on the surface of neutron stars, 
making this analysis of some interest in astrophysics.)
Moreover, the binding enhancement achieved by making the
system effectively one-dimensional can only account for
small effects, such as the fact \cite{AHS} that singly negative ions
always have infinitely many bound states in a magnetic field.
It cannot account for the binding of an additional $Z$ electrons.

The results in \cite{BR1} suggest that the primary mechanism for
binding additional electrons in strong fields is the fact that
the effective reduction in the strength of the electron-electron 
repulsion permits two electrons to be near the nucleus
simultaneously.  However, the one-dimensional confinement also
delocalizes the electron.  This effect is seen in the 
 Hamiltonian $h(N,Z,B^{-1/2})$ given by (\ref{eq:1dimHam}) 
where the effective mass is
$M = B^{-1/2}$ so that in strong fields the electrons behave
like extremely light  particles.  The uncertainty principle
then implies that trial wave functions which localize the
electrons cannot yield bound states.

Since Lieb's strategy \cite{Lb} for finding an upper bound on
$N_{\max}(Z)$ does not require an explicit localization,
it might seem well-suited to atoms in strong magnetic fields.
However, Lieb's method  actually has an implicit
localization (which is based on an idea
of Benguria \cite{Ben} for spherically symmetric atoms)
for which the localization error is zero in three dimensions.
However, as explained in \cite{BR1}, the localization error
is necessarily non-zero in one-dimension.  (This is a 
consequence of the fact that non-positive
potentials always have at least one bound state in one dimension.
Thus, the phenomenon of enhanced binding in one dimension actually
contributes to the delocalization of the electrons.) Using 
Lieb's method for the zero model, we were only able to show in
\cite{BR1} that $N_{\max}(Z,B) < 2Z + 1 + c \sqrt{B}$ for an
explicit constant $c$. In the interesting case $B = O(Z^3)$, this 
yields a bound of the form $N_{\max} < 2Z+ c Z^{3/2}$, rather than  a
linear one.

Surprisingly, one can get a better bound using the Ruskai-Sigal
localization method.  (See \cite{Rusk} for a summary.)
For both the zero model and the Slater model, we can prove
the following result.
\begin{theorem}  Let $N_{\max}(Z,B)$ be the maximum number of
electrons for which the Hamiltonian (\ref{eq:1dimHam}) has
a bound state, and assume that the potentials $\tilde{V}$ and
$\tilde{W}$ have either the form (\ref{eq:zero}) corresponding to
the zero model or the form (\ref{eq:V.slater}) and (\ref{eq:W.slater})
corresponding to the Slater model.  Then for every $\alpha > 0$ and
$\beta > 0$ there is a constant $C_{\alpha \beta}$ such that
\begin{equation}
 N_{\max}(Z,B) < 
    C_{\alpha \beta} Z^{1 + \alpha} B^{\beta}
\end{equation} where $\alpha, \beta$ can be arbitrarily small and, in the
case of the Slater model  $B \geq Z^{3 + \gamma}$ for
some $\gamma > 0$.
\end{theorem}
This result can be improved slightly to
\begin{equation}
 N_{\max}(Z,B) <  C_{ \omega}
    \left[ Z (\log Z)^2 + Z \log Z (\log B)^{1 + \omega} \right]
\end{equation}
where, as above, $\omega > 0$ can be arbitrarily small and
in the case of the Slater model $B \geq Z^{3 + \gamma}$.
Because the electrons in the one-dimensional model are
essentially bosonic, this is the best that one can hope
to achieve with the Ruskai-Sigal method.

In the case of the Slater model, the Landau level portion of the
wave function $\Upsilon$ is antisymmetric.  Hence, the one-dimensional
part of the wave function $\psi$ must be symmetric.  Even for
the zero model, it is physically reasonable to treat the electrons
in the one-dimensional model as essentially bosonic.  In the
Ruskai-Sigal method, the system is divided into a small ``inner'' ball
in which binding is precluded because the electrons are confined
to a small region, and an ``outer'' ball in which the localization
error becomes negligible as $Z \rightarrow \infty$.  For bosonic
systems, one can always squeeze the electrons closer together,
yielding a smaller cut-off $\rho$ than for fermions.  This feature
is the {\em only} factor which precludes extending the
proof of asymptotic neutrality in  \cite{LSST} to bosonic atoms.
This demonstrates that the localization error is not simply a
technical artifact, but a reflection of a real physical effect.

For atoms in strong magnetic fields, the cut-off radius $\rho$
is not small.  Instead 
$\rho \sim N \sqrt{B} Z^{-2} (\log \frac{Z^2}{B})^{-2}$ which
grows with $B$.  For $B = O(Z^3)$ and $N = O(Z)$, roughly
speaking (i.e., ignoring the log term) 
$\rho \sim Z^{1/2} \sim B^{1/6}$.
Thus, the localization method can be used to obtain a (non-optimal)
upper bound on $N_{\max}$ despite the fact that the electrons
are highly delocalized and the size of the ``inner'' region
becomes arbitrarily large as $B \rightarrow \infty$.  However, the
non-optimal bounds above are probably the best one can expect
from configuration space localization.  It seems likely
that a proof of better upper bounds will require the use of
phase space localization techniques.

\section{Properties of ${\mathbf V_m(x)}$}  \label{sect:Vm.props}

The functions $V_m(x)$ are even and well-defined for $x > 0$.
Although the primary interest in physical applications is for
integer $m \geq 0$, it is easy to
see from the form (\ref{vmequiv2})  that they can be
extended to complex $m$ with $\Re (m) > - 1 $.  For $\Re (m) > - \half $
they are also well-defined for $x = 0$.  In this note we
restrict ourselves to non-negative $x$ and  real $m > -1$.

It is convenient to define $V_{-1}(x) = \frac{1}{|x|}$ and note
that this is justified in the sense that
$\displaystyle{\lim_{m \rightarrow -1^+} x^{-1} V_m(x) = 1}$
for all $x > 0$.

We now summarize the properties of $V_m(x)$ for $x \in (0,\infty)$.
Unless otherwise stated $m$ is real and $m > - 1$.  For proofs
and further discussion, see  \cite{RW}.

\bigskip

\noindent{\bf Summary of Properties of $V_m(x)$:}
\begin{enumerate}
    \renewcommand{\labelenumi}{\theenumi}
    \renewcommand{\theenumi}{\alph{enumi})}

  \item   $V_m(x)$ satisfies the inequality
$\ds{\frac{1}{\sqrt{x^2 +m}}  > V_m(x)   > \frac{1} {\sqrt{x^2 + m + 1}}}$
where the upper bound holds for  $m > 0$ and the
lower for $m > -1.$

\item  $V_m(x)$ is decreasing in $m$.  In particular,
 $\ds{ V_{m+1}(x) < V_m(x) < \frac{1}{x} }$.

 \item The expression $ m V_m(x)$ is increasing in $m > -1$.

 \item For $m > -1/2$, the definition of $V_m(x)$ can be
extended to $x = 0$ and
\begin{eqnarray*}\label{vm0}
V_m(0) = \frac{\Gamma(m+\half)}{\Gamma(m+1)}.
\end{eqnarray*}
For integer $m$, this becomes
\begin{eqnarray*}
V_m(0) = \frac{(2m)!}{2^{2m}(m!)^2} \sqrt{\pi} = \frac
  {1 \cdot 3 \cdot 5 \ldots (2m-1)}{2 \cdot 4 \cdot 6 \ldots (2m)} \sqrt{\pi}
\end{eqnarray*}
while for large $m$ Stirling's formula implies
\begin{eqnarray*}\label{vm0.asymp}
   V_m(0)  \approx \left( \frac{m-\half}{m} \right) ^m
     \left( \frac{e}{m} \right) ^{1/2} \approx \frac{1}{\sqrt{m}}
\end{eqnarray*}
which is consistent with property (a). 

  \item  For all $m \geq 0$, $V_m$ satisfies the differential equation
\begin{eqnarray*}
  V_m^{\prime}(x) = 2x \left( V_m - V_{m-1} \right) .
\end{eqnarray*}
 \item  For each fixed $m \geq 0$, $V_m(x)$ is  decreasing in $x$.

 \item  For $a > 0$, the expression $aV_m(ax)$ increases with $a$.  Hence
 $aV_m(ax) > V(x)$ when $a > 1$ and $aV_m(ax) < V(x)$ when $a < 1$.

 \item   $V_0(x)$ is convex in $x > 0$; however, $V_m(x)$ is {\em not}
convex when $m > \half$.

 \item   For integer $m$, $1/V_m(x)$ is convex in $x > 0.$

\item  For integer $m$, the ratio $V_{m+1}(x)/V_m(x)$ is
 increasing in $x > 0.$

\item  The Fourier transform is given by
\begin{eqnarray*}
\widehat{V}_m (\xi ) \equiv 
\frac{1}{\sqrt{2\pi}} \int_{-\infty}^{\infty} V_m(x) \, e^{-ix\xi} dx = 
    \frac{4^{m+1}}{\sqrt{2 \pi}}\int_0^{\infty}
        \frac{s^m e^{-s} }{(|\xi|^2+4 s)^{m+1}}  ds
\end{eqnarray*}

  \item  For large $x$,   it follows from property (a) that
\begin{eqnarray*}
\frac{m}{2 (x^2+m)^{3/2}}
\leq  \frac{1}{x} - V_m(x) < \frac{m+1}{2x^3}
\end{eqnarray*}
while  (\ref{vmequiv2}) yields the asymptotic expansion
\begin{eqnarray*}
 V_m(x) = \frac{1}{x} - \frac{m+1}{2x^3} + \frac{3(m+2)(m+1)}{8x^5}
      +  O \left(\frac{1}{x^7} \right).
\end{eqnarray*}

 \end{enumerate}

\bigskip

The lower bound in (a) was proved earlier (at least for integer $m$)
by Avron, Herbst and Simon \cite{AHS}.
Properties (b) and (c) imply that
 $V_m(x)$ is decreasing in $m$, while $m V_m(x)$ is increasing;
this gives an indication of the delicate behavior of $V_m$.
The differential equation (e) can  be verified using 
integration by parts in (\ref{vmequiv2}).  Property (f)
 follows directly from  (b) and (e).
Property (g) follows  from  
(\ref{vmequiv2}) and the observation that
$\ds{ \frac{a}{\sqrt{a^2x^2 + u}}}$
is increasing in $a$.
It is useful in analyzing $ \tilde{W}(u-v)$ since it implies 
$\frac{1}{\sqrt{2}} 
              V_m\left(\frac{|u-v|}{\sqrt{2}}\right) < V_m(|u-v|).$
Property (h) follows from a straightforward analysis of the differential
equation (e) which implies that ${V_m^{\prime}(0) = 0}$ for
$m > \half$.  In the next
section, we will see that  the cusp at $x = 0$ and
the convexity of $V_m(x)$ in $ x > 0$ return when $V_m(x)$ is
replaced by $V_{\av}^N(x)$ as in the Slater model.

The convexity of $1/V_m(x)$ can be rewritten as
$$   \frac{1}{ \half V_m \left( \frac{x+y}{2} \right) } \leq
      \frac{1}{V_m(x)} +  \frac{1}{V_m(y)} $$
Using property (g) with $a = \half$, one easily finds that the
 this implies
\begin{eqnarray*}
   \frac{1}{ V_m(x+y) } \leq
      \frac{1}{V_m(x)} +  \frac{1}{V_m(y)}.
\end{eqnarray*}
Since $1/V_m(x) \approx |x|$ for large $|x|$, this subadditivity
inequality plays the role of the triangle inequality in applications.  
The proof of (i)  is extremely delicate.  Because $1/V_m(x) \approx x$
for large $x$, we need to prove the convexity of a function that is
nearly linear so that its second derivative is extremely close to
zero.  Proving that this derivative is positive is equivalent to
proving some rather sharp inequalities on the ratio
$V_m(x)/V_{m-1}(x)$.  

In the special case $m = 0$ these inequalities (which are 
also discussed in \cite{BR1} and \cite{SW}) are equivalent to
\begin{eqnarray}\label{opt.ineq}
          g_{\pi}(x) \leq  V_0(x) < g_4(x)
 \end{eqnarray}
for $ x > 0 $, where 
\begin{eqnarray} g_k(x) = \frac{k}{(k-1)x + \sqrt{x^2+k}}.
\end{eqnarray}
Multiplying (\ref{opt.ineq}) by 
$\displaystyle{ x = \frac{1}{V_{-1}(x)}}$ converts
this to a bound on the ratio $\displaystyle \frac{ V_0(x)}{V_{-1}(x)}$.
To obtain general ratio bounds, define
\begin{eqnarray}\label{def:Gkm}
    G_k^m(y) = \frac{ky}{ (k-1)y - m + \sqrt{(y+m)^2 + ky} }.
\end{eqnarray}
Then it is shown in \cite{RW} that
\begin{eqnarray}\label{eq:rat.bnd}
G_8^{m-1}(x^2) <   \frac{V_m(x)}{V_{m-1}(x)}  < G_4^m(x^2)
\end{eqnarray}
for all integer $m \geq 0$ and $x > 0 $.  The sense in which
these bounds are optimal is discussed in \cite{RW}.
 Our proof of
these inequalities relies on an inductive argument and, hence,
is valid only for integer $m$.  
A proof  extending them to general real $m > -1$ 
would immediately imply that properties (i) and (j) also
hold for  general real $m > -1$.

Another interesting open question is whether or not
$V_m(x)$ is convex in $m$?  In particular, is
$2V_m(x) \leq V_{m+1}(x) + V_{m-1}(x)$?  It follows from 
property (e) that this is equivalent to
asking if $V_m^{\prime}(x)$ is increasing in $m$.

\section{Recursion and Averaged Potentials}\label{sect:Vm.recurs}

Using integration by parts on (\ref{vmequiv2}) one easily finds that
$V_m$ satisfies the recursion relation
\begin{equation}\label{vmrecurs}
  V_m(x) = \frac{1}{m}
   \left[ (m- \half -x^2 )V_{m-1} (x)+ x^2 V_{m-2} (x) \right].
\end{equation}
 for all $m \in {\bf R}$, $m \geq 1$.  Iterating this, one finds
that when $m$ is a positive integer
\begin{eqnarray}\label{vpcor1}
 V_m(x) = \frac{1}{2m} \left[ (1 - 2x^2) V_{m-1}(x)  +
   \sum_{k=0}^{m-2} V_k(x) + 2|x| \right].
\end{eqnarray}

These relations are useful for studying $V_{\av}^N(x)$.  For 
example, it follows immediately from (\ref{vpcor1}) that
\begin{equation}
 V_{\av}^N(x)  = 2 V_N(x) - \frac{2x^2}{N}
   \left[ V_{-1}(x) - V_{N-1}(x) \right].
\end{equation}
This can then be used to show that ${V_{\av}^N(x)}$ is convex for all 
$x >0$.  Furthermore
$\ds{\lim_{x \rightarrow 0+} \frac{d~}{dx} V_{\av}^N(x) = 
- \frac{2}{N} }$, verifying that $V_{\av}^N(x)$ has a cusp at $x = 0$.

It is interesting to note that (\ref{vpcor1}) also implies that
there are polynomials $P_m(y)$ and $Q_m(y)$ of degree $m$
such that for integer $m \geq 1$
\begin{equation}
V_m(x)  =  P_m(x^2) V_0(x) + xQ_{m-1}(x^2).
\end{equation}
 These polynomials have many interesting properties.  In \cite{RW} it is
shown that
\begin{equation}
 P_m(y) = \frac{1}{m~B(m,\half)}~ e^{-y}
  ~_1F_1\left(\half,\half-m,y \right)
\end{equation}
where  $B(m,n)$ denotes the beta function and $~_1F_1(\alpha,\gamma,y)$
denotes the indicated confluent hypergeometric function.

\bibliographystyle{amsalpha}

\end{document}